# Lunar semimonthly signal in cloud amount

Nikolay Pertsev[1], Peter Dalin[2,3]

[1]A. M. Obukhov Institute of Atmospheric Physics, Russian Academy of Sciences, Pyzhevskiy per., 3, Moscow, 119017, Russia

[2]Swedish Institute of Space Physics, Box 812, SE-981 28 Kiruna, Sweden.

[3] Space Research Institute of Russian Academy of Sciences, Profsouznaya st. 84/32, Moscow, GSP-7, 117997, Russia.

**Abstract− Based on NASA satellite infrared and visible range measurements, cloud amount ISCCP_D1 summer nighttime data, representing the tropospheric cloud activity at Central Russia are examined over 1994-2007, and the lunar signal in the cloud amount was extracted. The ISCCP_D1 database was used to confirm previous results of Pertsev, Dalin and Romejko (2007) on the large importance of lunar declination effect compared to the lunar phase effect. Since this database provides much more information than it was used in that previous investigation, it has become possible to separate the lunar phase effect and the lunar declination effect in cloudiness. The relative cloud amount tends to grow with a change of lunar phase from a quadrature to the New Moon or Full Moon and with increasing of the lunar declination by absolute value. The both effects are statistically significant, the second one is a little stronger.**

**Introduction**

The lunar synodical semimonthly period has been found in precipitation (Adderley and Bowen, 1962), sunshine duration (Lund, 1965) and ice nucleus concentration in the troposphere (Bigg, 1963). Some more recent statistical results on this topic were mentioned in (Pasichnyk, 2002). The global temperature of lower troposphere, probably interrelated to the cloudiness, also provides the lunar synodical monthly and semimonthly peaks in its spectrum (Balling and Cerveny, 1995). The lunar synodical monthly and semimonthly periodicity was found also in the noctilucent clouds (Kropotkina and Shefov, 1976). Thus both tropospheric and mesospheric clouds are involved in some processes or process connected with synodical semimonthly oscillations, at least in restricted geographic regions in definite seasons of a year.

Dalin et al., (2006) have drawn an attention to mutual statistical dependence of the lunar phase and lunar declination. This mutual dependence is especially strong within the short seasonal segment (of many years), and must be taken into account in the statistical analysis, when we deal with clouds within definite seasonal segments. In (Pertsev et al., 2007) (we shall shortly refer to it as PDR07) en effort was made to distinguish between the lunar phase effect and lunar declination effect both in tropospheric nighttime summer cloudiness and noctilucent clouds over Moscow (Russia) region for the monthly and semimonthly periods, that led to a conclusion that the declination effect was significantly stronger for semimonthly lunar signal in the both types of clouds, whereas the lunar phase effect might arise artificially due to mutual statistical dependence of the lunar phase and lunar declination. The non-sufficient data number in the used in PDR07 database on both tropospheric and noctilucent clouds did not allow to come to a more precise result in that paper. As an index of tropospheric cloudiness, PDR07 used the rough estimations of the relative sky area, covered by tropospheric clouds, made by observers of noctilucent clouds within summer nighttime nights during many years. The main aim of the present paper is to check the results of PDR07 with more large and reliable database on cloudiness and to formulate a more precise conclusion on the influence of lunar phase and lunar declination on the cloud development.

**Modeling background on lunar tides in atmosphere**

While the lunar oscillations in the atmosphere may be carried by various physical mechanisms (Adderley and Bowen, 1962; Markson, 1971; Herman and Goldberg, 1978), the gravitational tide is usually recognized as most realistic or powerful mechanism. Several models were used to investigate the effects of lunar tides in the atmosphere (e.g. Chapman and Lindzen, 1970; Forbes, 1982). The lunar tidal forcing is described by a gravitational potential:

$$\Pi_L \approx -\frac{3}{2}\frac{gMr^2}{D^3}P_2(\cos q) \qquad (1)$$

where $g$ - the gravitational constant, $M$ - the mass of the Moon, $r$ - the distance from the Earth's center to a probe point at the Earth's surface or in the atmosphere, $D$ - the varying distance between the Moon and Earth, $q$ - the polar angle between the Moon's center and a probe point, $P_2(\cos q)$ is the zonal harmonic of degree 2 which is expressed as follows:

$$P_2(\cos q) = \frac{1}{2}[3(\sin^2 f - 1/3)(\sin^2 d - 1/3) - \sin 2f \sin 2d \cos(t-n) + \cos^2 f \cos^2 d \cos 2(t-n)] \qquad (2)$$

where $f$ is the latitude of a probe point, $d$ is the lunar declination (to the equator), $t$ is the mean solar local time in angular units and $n$ is the lunar phase angle which is equal to the difference between the longitude (i.e. difference in right ascensions in angular units) of the mean Moon and that of the mean Sun. Although expression (1) contains oscillations with different periods, in this paper we consider mainly semimonthly ones (~ 14 days). Such variations are described by the two components in (1). The first one, proportional to $\sin^2\delta$, is governed by lunar declination and has an average period of 13.66 days. The term $\sin^2 d - 1/3$ in (2) is varying from -0.33 up to -0.11 and has an average value -0.244, thus this oscillation has amplitude up to 67% of an average value of (1) at the given latitude *f*. The second semimonthly component is a product of the time-independent part of (2) and second harmonic of periodically (27.55 days) changing $D^{-3}$. The Fourier decomposition of $D^{-3}$ provides an amplitude of this second harmonic as small as 1% of an average value of (1). This simple description of semimonthly variations in the lunar tidal potential must be accompanied by the two important sophistications when one investigates the lunar semimonthly variations in geophysical data. The first one concerns a special data sampling: when the analyzed geophysical data are taken for the time close to constant local solar time, variation ~ *cos 2 (t- v)* with semidiurnal (12 h 25 m) period looks like a process determined by the double lunar phase 2v with semimonthly period 14.77 days (Chapman and Lindzen, 1970). The second sophistication arises from the statistical mutual dependence of lunar declination and lunar phase (Dalin et al., 2006), that exists in spite of the difference in average periods (27.32 and 29.55 days) of these variables.

**The used data on cloud amount and lunar position**

In this paper the ISCCP-D1 data (Rossow et al., 1999) of cloud amount were used. The cloud estimations are based on radiance measurements from the several satellites in visible and/or infrared range. From the very large volume of information, provided by ISCCP-D1 data, we used rather small part, seemed to be best comparable to the database, taken in PDR07. The relative cloud amount (RCA), i.e. a ratio of the number of cloudy pixels to the total number of pixels in a given grid cell, was used as a single index, representing a cloud amount. Only the summer data (from May, 16 to August, 16) of 14 years (1994-2007), for UT=21±1.5 h and from only the three grid cells (6010-6012) with latitude 56.25°±1.25°N and longitudes varying from 40.5° to 54°E, were put into the present analysis. Each of those cells has an area of $8 \cdot 10^4$ km$^2$ and located deep inside the continent to the east of Moscow (central part of Russia). The local solar time (LT), corresponding to the selected UT and the grid cells, can vary between 21h 36m and 02h 12m.

Orbital elements of the Moon motion (phase angle and lunar declination to the equator) were calculated from the fundamentals of celestial mechanics (Montenbruck and Pfleger, 2000).

**Statistics: cloud amount versus lunar position.**

The two dependencies, found in PDR07 for the lunar semimonthly oscillation in cloudiness, namely the dependence on lunar phase (cloudiness vs. cos *2v*) and the dependence on absolute value of lunar declination, reveal themselves in the ISCCP-D1 database as well. They are shown by the linear regression lines in Figures 1 and 2. Instead of non-uniformly distributed variable cos *2v*, we use a definition of the lunar phase $\Phi = \frac{1}{2}\arccos(\cos 2n)$, which is not distinguished between the lunar and anti-lunar points of the sky, so that the lunar phase is allowed to vary from 0 (New Moon or Full Moon) to 90° (First or Last Quarter).

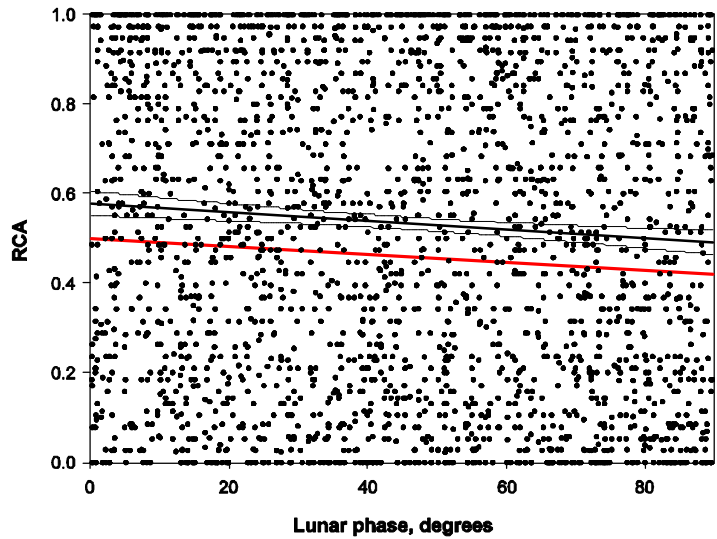

**Fig. 1. Lunar phases (defined by (3)) vs. RCA index. Each date (UTC=21 h) is represented by the three dots, corresponding to the three longitudinal cells. The linear regression dependence and its 95% confidence limits are marked by the black lines. The red line give the regression dependence for the case of removed data with RCA>0.975.**

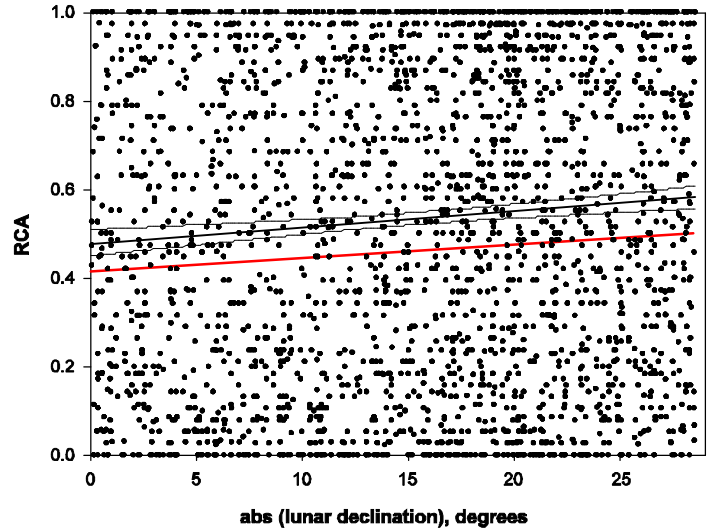

**Fig. 2. Same as in Fig. 1, but for absolute values of the lunar declination vs. RCA index.**

The 95% confidence intervals are also given in Figures 1 and 2 and show that the both regression lines are significant with more than 95% probability. The alternative semimonthly synodical harmonic $(\sim \arccos(\sin 2n))$ was also allowed to be extracted from RCA, however, as well as in PDR07, the latter dependence appeared to be very weak and not significant for any acceptable probability.

The statistical dependence between the lunar phase and lunar declination (Dalin et al., 2006) must be further taken into account. The statistics in the previous studies (Dalin et al., 2006; PDR07) did not allow us to consider the simultaneous influence of the lunar declination and lunar phase. The number of dots in the present statistics is large enough (2592) to separate the two effects by using bilinear least-square fitting: $RCA = const + b_1 \cdot \Phi + b_2 \cdot |d|$

The more definite (less scattered) is relationship between the two arguments, the larger are the errors of the regression coefficients $\beta_1$ and $\beta_2$. If there was a substantially less-scattered dependence between the lunar phase and absolute values of the lunar declination, the separation of the two effects would not be possible. Fortunately, the scattering of the two arguments around their mutual regression appears to be large enough to obtain not too large errors in the regression coefficients. In the aftermath, the regression coefficients are found to be as follows: $\beta_1$=(-5±3)·$10^{-4}$ degr$^{-1}$, $\beta_2$=(2.6±1.1)·$10^{-3}$ degr$^{-1}$. Relative contributions of the two dependencies to the RCA standard deviation are as large as 3.5% for the lunar phase, 6% for the lunar declination and 8.5% for the both ones. Therefore, we confirm the previous finding (PDR07) on the lunar signal in RCA: the lunar declination effect in summer nighttime cloudiness is stronger than the lunar phase effect, but now we provide more precise conclusion that the both of them are statistically justified and give their own contribution into the cloud variability.

**Discussion and conclusion**

The main question to be discussed is a separation of the lunar semidiurnal and the lunar semimonthly oscillation in spite of mentioned aliasing between them. Since the cloudiness data are taken for LT close to midnight, then $t \approx 0 + 2\pi n$, so for such data sampling the semidiurnal oscillation ~ cos *2 (t- v)* behaves as semimonthly ~ cos 2*v*. But the gravitational potential (1) contains no semimonthly terms, governed by lunar phase *v*. That is why we assume that the found lunar phase effect describes the lunar semidiurnal oscillation. The insignificance found for odd effects in the lunar phase (~sin 2*v*) confirms such a suggestion. Just this situation must occur according to (1, 2).

As to the declination effect, it must be semimonthly tide, because it corresponds to the only term in (2), which is governed by variations in $\delta$ and which is separated from $v$. Besides extracting of signal, proportional to $|\delta|$, an extraction of other semimonthly changing variables such as $d^2, \sin^2 d, \frac{\sin^2 d}{D^3}$ was made. The difference in fitting quality, measured by standard deviation of residual series, appeared to be miserable between all of those variables. Thus the found dependence of RCA on $|\delta|$ describes likely a ~ $\frac{\sin^2 d}{D^3}$ semimonthly oscillation of the gravitational potential (1).

The results described in this paper are obtained under the spatial and temporal restrictions on analyzed data (very narrow ranges of LT, latitudes, longitudes and limited season). On the one hand, it allows us to avoid the additional statistical noise caused by diurnal, seasonal and geographical variations in cloud formation and latitudinal variations in lunar perturbations (it is not clear, whether the lunar effect would show up the similar dependences in different types of clouds, in different seasons and latitudes). On the other hand, it is important at the first stage to prove the existence of the distinct lunar declination signal in the cloudiness for the more detailed studies of lunar perturbations in clouds in the next stages. Much more complicated analysis should be made, taking into account different latitudes, longitudes, LTs, seasons, continents/ oceans and allowing the break of even symmetry both in lunar phase and in declination (for other LTs the ~$\sin 2\nu$ term in (2) would appear, and in other seasons the mutual statistical dependence of the lunar phase and declination becomes more complicated).

Returning to the first papers devoted to the lunar synodical effect in variables closely related to the precipitations and the tropospheric cloudiness (Adderley and Bowen, 1962; Bradley et al., 1962; Bigg, 1963; Lund, 1965), one can find that their results are hardly comparable with ones described in the present paper , because they did not consider the lunar declination, besides their data were not fixed to LT or uniform in LT, e.g. Bigg (1963) considered daily mean ice nucleus concentration and daily mean rainfall measure with unknown LT-distribution. The latter does not permit to separate the lunar semidiurnal and lunar semimonthly oscillation.

Our conclusions are as follows:

1. For the first time the lunar signals in cloudiness have been extracted from a reliable database.

2. These lunar signals contain the semimonthly tide governed by the variations in the lunar declination and the semidiurnal tide, governed by the changes in the lunar phase (under the condition of the fixed solar local time).

3. The extracted lunar signals seem to fit the theory of the lunar gravitational tide.

**Acknowledgements.** The ISCCP-D1 data of cloud amount were obtained from the NASA Langley Research Center Atmospheric Science Data Center.